# The role of physiological complexity changes in resting-state EEG in clinical effectiveness of rTMS and tDCS in treatments of resistant depression

*Milena Čukić[1,2], PhD*

[1]Department for General Physiology and Biophysics, University of Belgrade, Belgrade, Serbia
[2]Amsterdam Health and Technology Institute, HealthInc, Amsterdam, the Netherlands

## Abstract

The present literature about possible mechanisms behind the effectivity of noninvasive electromagnetic stimulation in major depressive disorder (MDD) is not very rich. Despite extensive research in applications for clinical practice, the exact effects are yet not clear. We are comparing our previous results about the complexity changes induced by TMS, and tDCS which are known to modulate neural dynamics. Also, we are reviewing different biomarkers of complexity changes connected to depression, and how they change with the stimulation. TDCS is low-intensity TES, known to have polarity specific effects (neuromodulatory effects), and rTMS is inducing an electric field in the tissue circumstantially via Faraday's law. Both nonlinear modalities of electromagnetic stimulation may affect the levels of physiological complexity in the brain. We also compare the changes of complexity in electroencephalogram (EEG) and electrocardiogram (ECG), as potential future predictors of therapy outcome.

## Introduction

The course of depression is very often recurrent and can become chronic with highly likely relapse rates within one year of remission from 35% to 80% (Eaton et al., 2008; Fekadu et al.,

2009). Usual treatments for depression are pharmacological and psychological, but Rush and colleagues report that one third fail to reach remission (Rush et al., 2006). Non-invasive electromagnetic modalities of stimulation, like transcranial magnetic stimulation (TMS) and transcranial electric stimulation (TES) are alternatives to this usual practice. We are going to briefly review several application modalities of TMS and low-intensity modality of TES, namely tDCS. Here we want to focus on elucidating the effectiveness of two electromagnetic treatment considered 'novel.' My putting it under the quotation marks is intentional because transcranial magnetic stimulation was launched as a commercial stimulator 1985 (Prof. Antony Baker, 1985) and it cannot be still novel after almost 35 years of research and diagnostic applications. Similar reasoning applies to tDCS which is low-intensity stimulation modality of transcranial electrical stimulation (TES), but still, some details important for an understanding of its mechanisms are missing. Like Opitz put it in recent research, the important point is in interpretability of TES effects (Opitz et al., 2018): 'if electric fields are delivered inconsistently, but effects are observed nevertheless, the results are more difficult to interpret because effect could be driven by other incidentally affected brain regions.'

TMS was initially introduced as a non-invasive tool for investigating and mapping cortical functioning and connectivity (Barker et al., 1985). TMS primarily use a strong magnetic field to induce an electric field in the surface layers of the brain (cortex) and without pain characteristic to ECT, initiate optimally focused activation of neural structures. One of its modalities used in psychiatry is repetitive TMS (rTMS). While standard TMS induce single (or sometimes) paired pulses, repetitive TMS delivers repeated pulses showed to enable more extended modulation of neural activity. It can be slow (if repetitions are up to 1Hz/one per second) or fast (with 10Hz or higher frequencies). Fast rTMS have excitatory, while slow rTMS applications have inhibitory effects (Rosa and Lisanby, 2012). It is interesting to note that rTMS was finally confirmed for reimbursements in a treatment-resisting depression just last year (2018) in the Netherlands, 34 years from its début in medicine (US Food and Drug Administration- FDA approved the first rTMS device for treatment of MDD in 2008). The basis for rTMS treatment is a poor response to at least one pharmacological agent in the current episode (O'Reardon et al., 2007). We are going to elucidate on that question 'how long it takes for a scientifically proven innovation to be translated to clinical practice?' The scientific literature about a safe and useful application of TMS is so rich that it is hard to believe that it took for so long. The reason for using TMS in treatments of depression came from a demonstration of functional impairments in prefrontal cortices and limbic regions (Atkinson et al., 2014) and very well based frontal asymmetry in alpha (Allen et al., 2004). There are several different modalities of rTMS use. Since high-frequency stimulation can be uncomfortable during the initial stimulation period, low-frequency rTMS was introduced to minimize headaches and scalp discomfort; it is also lowering the risk for developing seizures (Rossi et al., 2009). When the stimulation is ongoing over the left and right dorsolateral prefrontal cortex-DLPFC (simultaneously, or stimulation one side after another), it is called Bilateral application of rTMS. Conca and colleagues argue that this kind of application may reinstate any imbalance in prefrontal neural activity (Conca et al., 2002). Fitzgerald showed that likelihood for a clinical response increases by providing both types of stimulation (Fitzgerald et al., 2006). Deep TMS (dTMS) can stimulate larger brain volumes and deeper structures (Roth et al., 2007) which is maybe more relevant to the pathophysiology of depression (Costafreda et al., 20013; Kwaasteniet et al., 2013; Atkinson et al., 2014). Theta burst stimulation (TBS) (Huang et al., 2005) utilize high and low frequencies in the same stimulation train; it delivers three bursts at high frequency (50Hz) with an inter-burst interval of 5Hz (theta range) frequency. Two different approaches to the application of TBS exists: continuous and intermittent (cTBS and iTBS). cTBS delivers 300 to 600 pulses without interruption, and iTBS delivers 30 pulses (trains) every 10s during 190s (600 pulses in total) (Chuang et al., 2015). While cTBS reduces cortical excitability, iTBS increases it; they are mimicking the process of long term potentiation (LTP) and long term depression (LTD) (Stagg and Nitsche, 2016; Huang et al., 2005). LTP allows for modulation of synaptic strength that stabilizes for days, months, or even years and has therefore been postulated as a likely candidate for memory formation in the brain (Anderson and Lomo, 1966; Bliss and Lomo, 1973). LTD has been studied extensively in the hippocampus and refer to highly specific processes; synaptic plasticity with very similar properties has been demonstrated in the neocortex and is therefore commonly referred to as LTP-like plasticity (Stagg and Nitsche, 2016). The main advantage of TMS is the reduced time of treatment in comparison to all other rTMS modalities. For conventional rTMS it takes 20-45 min, while for TMS it is typically less than 5 min; also it operates at 80% of the resting motor threshold (MT) and may be more comfortable than rTMS. Another modality of stimulation synchronized TMS (sTMS), is a new treatment paradigm which delivers stimulation synchronized to an individual's alpha frequency (Jin and Phillips, 2014). It is based on rTMS ability to make entrainment of oscillatory activity to the programmed frequency of stimulation with the aim of restoring normal oscillatory activity (Leuchter et al., 2013). It does not cause neural depolarization; therefore it may cause fewer adverse effects.

In opposition to all modalities mentioned above of TMS, tDCS which is low-intensity modality of TES has specific advantage of use; its low cost and portability are opening the possibilities of use out of the clinical setup. tDCS is not triggering action potentials directly (unlike TMS), and it is shown that it modulates cortical excitability by shifting neural membrane resting potential (Nitsche et al., 2008). TDCS effects can outlast the stimulation period up to two hours (Nitsche, 2001) and even show significant effects for more extended periods. In our recent research, we showed that tDCS is able of causing the shift in the brain-state (Čukić et al., 2018) based on EEG recordings reconstruction in phase-space, for at least half an hour. The primary advantage of tDCS compared to any TMS is its ease of administration, portability, much less recorded adverse events (AE) and lower price. Mutz and colleagues aimed at examining the efficacy and acceptability of non-invasive brain stimulation in adult unipolar and bipolar depression (Mutz et al., 2018). They found the most substantial evidence for high-frequency rTMS over left DLPFC, followed by low-frequency rTMS over the right DLPFC and bilateral rTMS. Intermittent TBS provides a potential advance in terms of reduced treatment duration. Authors also stated that tDCS is a potential treatment for non-resistant depression which has demonstrated efficacy in terms of response as well as remission (Mutz et al., 2018). Tuomas Neuvonen and colleagues started making their anytime anywhere tDCS applications in Soma already. They elaborated on interindividual differences as one of the major sources of misinterpretation (Fonteneau et al., 2018)

**Meta-analyses on the efficacy of rTMS and tDCS in treatment-resistant depression**

In the recent tDCS review study, authors discussed the interaction of induced electric field (EF) with tissue, and on electroporation and concluded that the intensities needed for electroporation remain orders of magnitude above tDCS (Antal et al., 2017). Typical current in tDCS montage is 1-2 mA (0.03-2mA/cm$^2$ current to electrode area ratios depending on the electrode size) which results in cortical EF strengths up to 0.4-0.8 V/m (Ruffini et al., 2013b). Both the applied current and the resting brain EFs are ~1000 times lower than those for pulsed stimulation used in electroconvulsive therapy (Alam et al., 2016) and are considered to be below the required intensity for evoking an action potential in resting cell (Radman et al., 2009). They are still capable of modifying the spontaneous firing rates of neurons, and also of inducing plasticity processes in neural networks (Ranieri et al., 2012; Jackson et al., 2016). Neural networks generate their own characteristic EFs in intracellular spaces, and this kind of stimulus can influence their molecular and structural changes (Frohlich and McCormick, 2010). Based on the work of Mutanen and Ilmoniemi on the induced brain-shift in an application of TMS, we later aimed at elucidating the effect of tDCS by utilization of the same method. Frohlich and McCormick detected the alteration of transmembrane potential for 0.5-1.3 mV, with the intensity of the induced electric field of 2-4 mV/m. Both Miranda and Saturnino predicted values of the maximum EF in the cortex of realistic head model are between 0.2 V/m and 0.5 V/m (if the current is 1mA) (Miranda et al., 2013; Saturnino et al., 2015). According to Nelson and Nunneley, tDCS induced energy in the cortex is 0.1 mW/kg, for the current of 1mA, a conductivity of 0.4 S/m and the median value of EF 0.5 V/m (Nelson and Nunneley, 1998). Antal and other experts concluded in their 2017 guidelines that those values caused by tDCS are safe (Antal et al., 2017). Wagner reviewed similar details important to understand the electromagnetic interaction involved from previous literature (Wagner et al., 2007). In case of TMS, a magnetic field (~1-4T pulsed over 0.125-1ms dependent of parameters) induces an electric field that drives currents in the brain of a magnitude approximately $5.13 \times 10^{-8}$ A/m in the cortex per 1A/s steady-state source. Currents are carried by free charges and ions (known as ohmic or resistive currents, or volume conduction) or through the depolarization of dipoles in the tissue layers (Wagner et al., 2007).

Fregni and colleagues were searching for predictors of antidepressant response in clinical trials of transcranial magnetic stimulation (Fregni et al., 2006). At the time they did their research, it was already known that rTMS has a significant antidepressant effect, the results were heterogeneous. They hypothesized that individual patients' characteristics might contribute to such heterogeneity in a sample comprising of 195 patients. Results showed that age and treatment refractoriness were significant negative predictors of depression improvement when adjusted to other significant predictors. The findings of Mutz and colleagues (2018) are in line because they also found that younger people are more probable to be responders. Another study on the effectivity 'Transcranial direct current stimulation in treatment-resistant depression: a randomized double-blind, placebo-controlled study' was published in 2012 (Palm et al., 2012). They concluded that 'anodal tDCS, applied for two weeks, was not superior to placebo treatment in patients with treatment-resistant depression. However, secondary outcome measures are pointing to a positive effect of tDCS on emotions'. The authors recommended modified and improved tDCS protocols to be carried out in controlled pilot trials to develop tDCS towards an effective antidepressant intervention in therapy-resistant depression. Berlim and colleagues published two studies on the same topic in 2013 and 2014 (Berlim et al., 2013, 2014). The first one was about the efficiency of tDCS, and the other on the efficiency of rTMS. They first concluded that the clinical utility of tDCS as a treatment for MDD remains unclear when clinically relevant outcomes such as response and remission rates are considered.

Moreover, they recommended future studies on larger samples. In their 2014 study, they concluded that 'Meta-analyses have shown that high-frequency (HF) repetitive transcranial magnetic stimulation (rTMS) has antidepressant properties when compared with sham rTMS.' However, its overall response and remission rates in major depression (MD) remain unclear (Berlim et al., 2014). Bennabi and his colleagues concluded in their 2015 study that 'tDCS did not induce the clinically relevant antidepressant effect in active and sham stimulation groups. There was no impact on psychomotor and neuropsychological functioning' (Bennabi et al., 2015). So, as Fregni pointed out, the results are still heterogeneous.

Brunoni and colleagues aimed to assess tDCS efficacy and to explore individual response predictors in their 2016 study (Brunoni et al., 2016). Since tDCS was extensively investigated for the treatments, particularly of major depression, they performed another review study by utilization of individual patient data (contrary to aggregate data approach in many review studies before). The current literature confirms that tDCS can induce neuromodulatory changes in cortical activity, and essential for depression, to ameliorate depressive symptoms (Brunoni et al., 2012). Its antidepressant effects are based on previous findings of hypoactivity of the left dorsolateral prefrontal cortex (DLPFC). Brunoni concentrated on providing precise estimates of tDCS efficacy based on depression improvement, response and remission rates. They tried to identify variables associated with tDCS efficacy (Brunoni et al., 2016). They also used several up-to-date software tools for detecting the bias, to re-analyze the original data and so on. After careful literature research and selection (among 153 studies), their review study comprises of six studies, only (they included studies which are double-blinded, sham-controlled, reporting on a number of repetitions, dosage, medication, and other detailed data). However, on end among those six studies were three (half of all selected) studies published by the same group which is also a bias (Loo et al., 2010; Loo et al., 2012; Brunoni et al., 2013). The primary dependent variable was clinical response since the authors claim that it has better performance over remission which presented heterogeneity of cut-offs. They used a number needed to treat (NNT) as a measurement which illustrates clinical intervention effectiveness; the higher the NNT, the less effective the intervention. The authors found significant effect sizes of the efficacy of active vs. sham tDCS in terms of depression improvement response and remission, with overall response and remission rates of active tDCS of 34% and 23% respectively (NNT of 7 and 9) (Brunoni et al., 2016). The authors further compared their own meta-analysis study (on six studies, half of which are their own) with a Cochrane meta-analysis assessing the efficacy of antidepressant drug treatments in primary care (Arrol et al., 2009) finding their results are in line.

Brunoni and colleagues also claim their findings to be in the same range as another meta-analysis for depression which utilized active vs. sham repetitive transcranial magnetic stimulation (rTMS) that found a response and remission NNTs of 6 and eight respectively. So, for treatment response and remission rTMS meta-analysis study(Berlim et al, 2014) found absolute rates of 29.3% and 18.6%, while tDCS meta-analysis found 34% and 23% respectively. The only problem here was that out of 289 patients included in tDCS study there were just 147 patients with an active stimulation, and in rTMS study, the sample is almost ten-fold; n=1371. From the point of statistical learning, the other sample would be expected to give us reliable results, while the first sample for this type of comparison is considered to be very modest.

Brunoni and colleagues also recommended two potential applications of tDCS in treatments for depression: in primary care settings and as a non-pharmacological, neuromodulatory therapy for depression (Brunoni et al., 2016). In another meta-analytical study performed by Antal et al. (2017) the authors commented on limitations of Brunoni study, and used a much larger sample, covering the publications and data collected from approximately 8000 people. They concluded that tDCS is considered safe, giving guidelines for avoiding Adverse Events (AE) as well as questionnaires for further research (designed to prevent all previously reported AEs).

In their meta-analysis comprising of 56 previously published studies, Mutz and colleagues found about actual effectivity of different modalities of TMS used for treatment-resistant depression (Mutz et al., 2018). They reported that high-frequency rTMS (over left DLPFC) was associated with improved rates of response as well as remission in comparison with sham treatment. Only one of the reviewed studies recruited bipolar depression patients (Nahas et al., 2003) but with demonstrated antidepressant efficacy. They also found that low-frequency rTMS over the right DLPFC was associated with significantly higher response and remission rates than sham stimulation. Low-frequency rTMS over lDLPFC did not show significant improvement (there were no significant differences in response rates compared to sham). Bilateral rTMS demonstrated significant improvement in response but not remission rates compared to sham. When dTMS is concerned, it also showed significant improvement (for both response and remission rates), but response rates were marginally higher relative to sham. Neither response nor remission rates for sTMS were significantly higher than a sham. Intermittent TBS over the left DLPFC was associated with a fivefold improvement in response rates compared to sham. No evidence was found of improvement in case of cTBS over rDLPFC and bilateral TBS. In case of tDCS, they confirmed significant improvement in both response and remission rates in comparison with sham; their analysis suggested tDCS to be effective only in patients with non-treatment resistant depression (Mutz et al., 2018). In contrast to most rTMS trials, tDCS showed a potential initial therapeutic option for depression.

### The evidence that rTMS and tDCS are activating deeper structures

A study performed by Li and colleagues provides a valuable insight into mechanisms by which tDCS may modulate cognitive function and also has implications for the design of future stimulation studies (Li et al., 2018). They showed that the effects of tDCS on brain activity are dependent on the cognitive state. Both anodal and cathodal tDCS produced widespread BOLD changes in brain areas anatomically remote from the critical area being stimulated. Those results are in line with Sven Bestmann's findings, which first demonstrated that even with the sub-threshold stimulation of TMS, deep structures are activated (Bestmann et al., 2003 and 2004). It may come as a surprise, but we are looking here at the evidence that both tDCS and rTMS are affecting other structures than intended. It may be that it is in connection with the finding that severe depression does have the problem with functional connectivity, due to the deep white matter tracts significant in a fronto-limbic system (Kwaasteniet et al., 2013; Kim et al., 2013). The focus of a majority of researchers was for so long at defining the polarity specific and other noticed effects, but the problem is that both methods are noninvasive and therefore only direct measurement can be a solution for this riddle.

Opitz et al (Opitz et al., 2015, 2016, 2018) and Alekseichuk (Alekseichuk et al., 2018) applied the robust physical approach in elucidating this complex interaction between the induced electrical field and brain tissue. Opitz performed the direct measurement from implanted electrodes (in direct empirical validation of TES) in patients who were operated in a procedure intended to implant a grid of stimulation electrodes for their epilepsy, and compared the results with the cohort of 25 healthy people (neurotypical individuals) from connectome project. The motivation for their work was the lack of exact measurements and the problem with variability between individuals. Opitz and colleagues combined direct intracranial measurements of electric fields generated by TES in surgical epilepsy patients with computational modeling (Opitz et al., 2018). They directly validated the computational models and identified key parameters needed for accurate model predictions. They also derived practical guidelines for a reliable application of TES in terms of the precision of electrode placement needed to achieve a desired electric field distribution. When comparing the measured and predicted values of induced electrical fields, they encounter the problem of conductivity constants values. They are reported in the literature with great variety, so further research is needed (Opitz et al., 2018). To establish standards in this field would be of significant importance to increase the reliability of future stimulation protocols. When they were testing the model developed for the study they found that maximum strength of the induced field predicted by the model was larger than that they directly measured. The mean electric field strength was 0.058 mV/mm for the measurement results, while the mean electric field strength over the best fitting stimulation results was 0.100 mV/mm (range 0.071-0.122 mV/mm) (Opitz et al., 2018). In another patient with a different implanted grid of electrodes, the trend was in the opposite direction; mean measured field strength was 0.115 mV/mm, while predicted was 0.060 mV/mm.

The authors concluded that computational models using current standard conductivities slightly misestimate the measured field estimate strength, suggesting the need for individual adjustments of conductivities (Huang et al., 2017). They also demonstrated the importance of factors like the accuracy of the skull or surgical component modeling; other factors as gyral folding and CSF thickness showed to have a profound effect on TES induced fields (Opitz et al., 2015). They urge other researchers to direct their efforts to directly measure conductivities to improve realistic modeling, or even include more tissue types (Aydin et al., 2014). It is based on their finding that the variation of the position of an electrode may maximally vary to 1cm, due to folding in the brain surface (electricity concentrate differently on sharp and flat surfaces, depending on geometry) in order to provide efficient stimulation. In their work, it is also confirmed how important it is to accurately model the shape of electrodes for realistic field calculations (Saturnino et al., 2015). In another study (Alekseichuk et al., 2018) they systematically evaluated the induced electrical fields and analyzed their relationship to brain anatomy (TMS and TES induced fields were compared in finite element method, in a mouse, capuchin monkey, and human model). The theoretical explanation and final calculation of the induced electrical field during the stimulation was necessary for those who are further developing computational models.

**Results from our previous research about the changes of complexity induced by the application of TMS and tDCS**

In my Doctoral dissertation, and before that in my magisterial thesis, I explored the parameters influencing the strength of stimulation (Čukić, 2006) and the changes in complexity due to TMS (Čukić, 2011). I am going to report here some of our results which can help understand the changes that both TMS and tDCS can induce, based on co-recording of the electrophysiological signal. Now, if we want to understand how a stimulus is acting upon a system, we can observe the levels of complexity before the stimulation, and after the stimulation and by comparing those levels we can conclude about the influence. This is precisely what we did with two mentioned modalities of stimulation. In our work from 2011/12, we compared single pulse TMS induced complexity changes with the complexity changes induced by peripheral electrical stimulation (the same target muscle was FDI), and we showed that single pulse TMS is able to decrease the complexity of the system (Čukić et al., 2013). In the first set of results, we aimed at examining the changes in surface electromyogram (EMG) recorded from the first dorsal interosseous muscle (FDI) which was the target muscle in TMS protocol. In that protocol, we were also changing the level of voluntary contraction to test how the amount of present force in the muscle can interfere with the effect of stimulation. On the first graph from those results (Figure 1) we can see surface EMG recorded in that experiment and also magnified epochs of that signal we used for further fractal analysis.

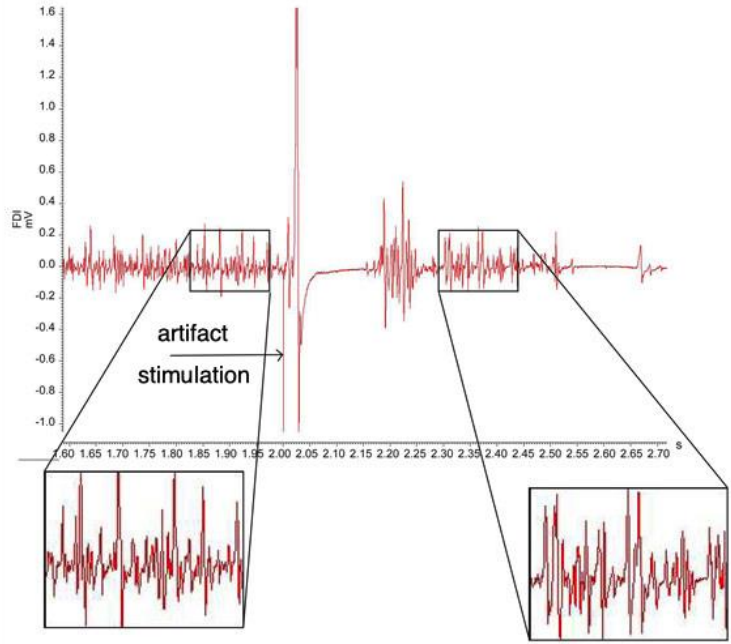

Figure 1: The raw electromyogram signal recorded from the target muscle (FDI) in a TMS stimulation protocol. The epochs for analysis extracted from the trace 'before' and 'after' the stimulation are magnified. They were used for further analysis to probe for the complexity changes induced by the stimulation (Čukić et al., 2013).

After comparative classical spectral and fractal analysis, we found the significant changes in the epochs recorded in participants (all healthy persons) before and after the stimulation. The following two illustrations are showing both the spectral and fractal result of the analysis.

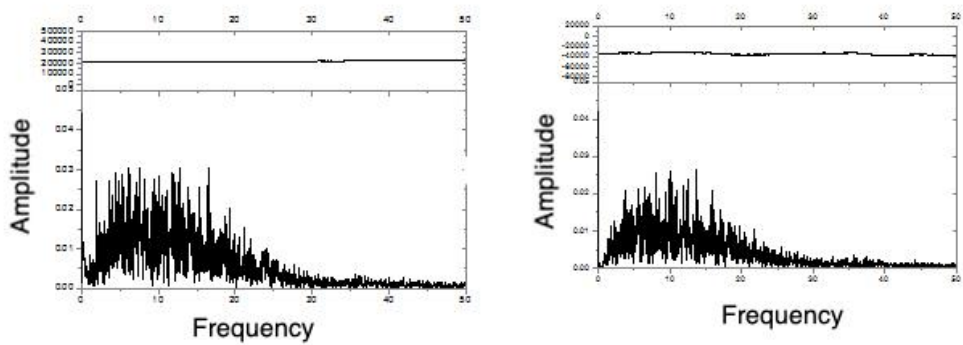

Figure 2: The distributions of frequnecies (spectra) before (left) and after (right) stimulation in the surface EMG signal; the change induced by the stimulus are not immediately obvious (Čukić et al., 2010).

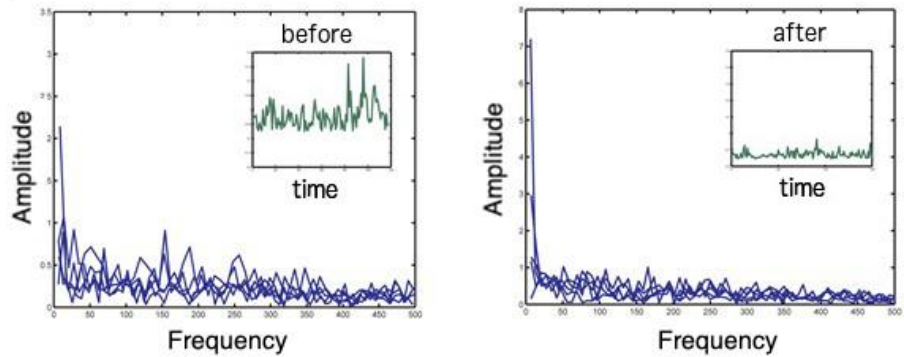

Figure 3: Amplitude spectra of surface EMG before (left) and after (right) the TMS stimulation. The difference in complexity levels can be seen with a naked eye (Ćukić et al., 2010).

It can be seen from both graphs that the drop in complexity of the signal is apparent. The level of complexity in a healthy system (being a brain, a heart or a muscle) is showing systems' ability to adapt. When an influence out of the dynamical system impact the system, the change can be observed via changes in complexity (or irregularity). In this case, it can be seen that after the application of TMS, the complexity of surface myogram reduced. One can see that on the following figures: 4 and 5. The first is showing a histogram, constructed from fractal dimensions calculated from sections of EMG signal, showing the distribution of its values. The red line compared to black dotted one is showing how those FDs calculated from epochs of EMG before and after the stimulation are distributed. Those representing 'after' sections are shorter, meaning that the facilitation happened during this stimulation. Another figure (4) is representing the mean values of calculated FD (from EMG) showing a significant difference between them.

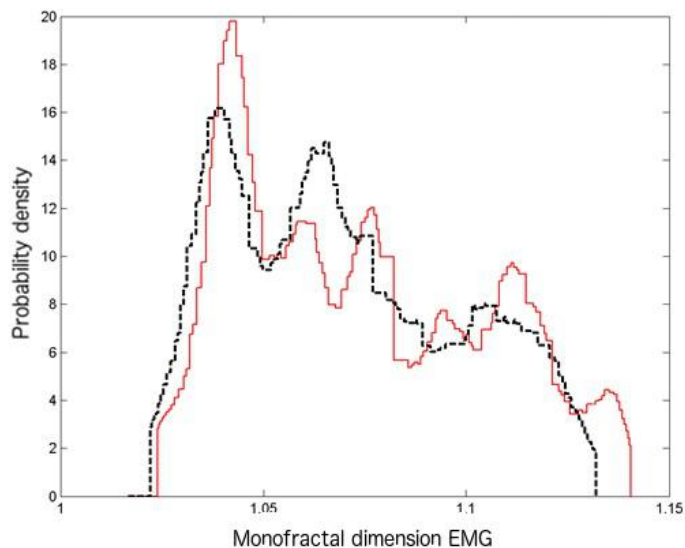

Figure 4: The changes in Higuchi fractal dimension calculated from surface EMG epochs before (red) and after the stimulation. The distribution is slightly shifted to the left, illustrating the shortened latencies in the response after the stimulation (Čukić, 2006).

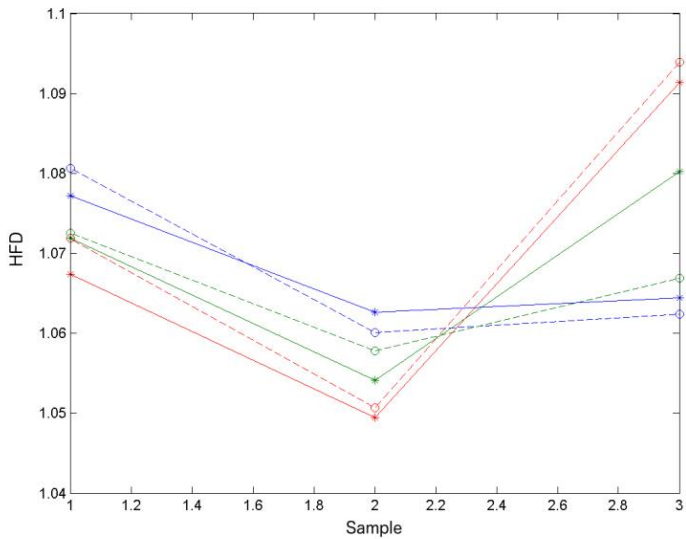

Figure 5: The average values of Higuchi Fractal dimension before (dotted line) and after the stimulation; it can be seen that the complexity of signal dropped after the stimulation (Čukić, 2012).

Several years after that experiment, our colleagues from Finland published a study which examined the change TMS induced in brain state (in phase-space) of 10 healthy participants (Mutanen et al., 2013). They applied recurrence plot analysis on EEG concurrently recorded with TMS, and they also stimulated motor cortex (the same target muscle as we did in our experiment). This is why I believe that our results are in a way comparable and compatible. They showed that even some time after the stimulation is over, the system (brain) stayed in a highly improbable and higher energy state in comparison to the resting state before the stimulation. The following diagram is illustrating the process they described in their work (Mutanen et al., 2013), on our graphical representation similar to theirs (Figure 6).

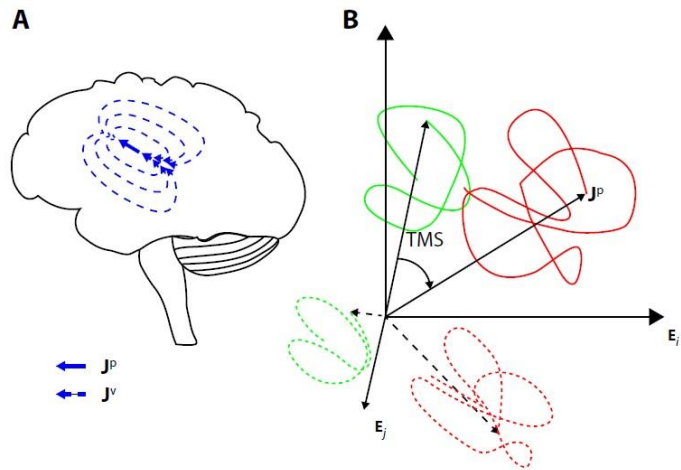

Figure 6: A) The blue arrow represents the primary current source, a flow of ions in synapses in the motor cortex. The dashed blue line represents returning volume current. Jp is the primary current density (induced by the stimulation). B) This schematic representation shows how TMS affects brain-state. The green and red curves are pre- and post-TMS trajectories (of a brain state). The spontaneous activity (green) occupy one region in state-space. After the stimulation, it occupies another region (red) which is in a higher energy state. Gradually, over time, the system will return to lower energy and more probable state. The projections of those two states are in EEG signal space, $i$ and $j$ standing for two different channels. In the signal, space trajectories are measured only at discrete time points (dotted projections, red and green).

The green line is representing the trajectory of a system before the stimulation, and red line the trajectory in another part of phase space after the TMS. Both trajectories have the projections on the plane (dots in the plane $E_i$ $E_j$), which are the samples of recorded EEG (Ilmoniemi and Kičić, 2010). $J_p$ and $J_v$ are the current densities illustrating the ongoing process during the stimulation; $J_p$ is causing the movement of charges in the tissue (it appears due to the induced electrical field in the tissue), and $J_v$ is volume conductor density current which is a counterpart process to the first one. In essence, they showed that due to the stimulation the system is staying in elevated state (the brain-shift is confirmed by the measures of Global recurrence, MSS and SV) long enough so we can detect it.

Next thing we did, was to test the same methodology, but on another type of stimulation, tDCS. We are presenting here part of the results which are dealing with complexity changes. On figure 6, it can be seen how HFD and SampEn detected the changes in complexity after tDCS stimulation. We know that this kind of stimulation is very different, due to the much lower level of energy delivered to the tissue, and we know that it is considered to have just modulatory character. In this research (in 2015) we tried to test the changes in complexity before and after the stimulation ('before' was considered a resting state since we did not have sham to compare to, and after was immediately after-$t_1$, or a half an hour after-$t_2$). However, if you can detect changes that survive more than half an hour later, does that imply early plastic-like changes? Unfortunately, we were not able to confidently interpret the results, because contrary to our previous knowledge about just modulatory effect of tDCS we detected temporary decrease and after that increase of complexity measured by Higuchi fractal dimension and Sample entropy.

What is more, it looked as at some combinations there is no difference between cathodal and anodal stimulation, in terms of complexity changes. Due to the small sample, we opted to apply another methodology set, but are working on repeating with the same method in presently ongoing research. Nitsche showed much earlier that induced changes in system survive more than two hours (Nitsche, 2001). It would be interesting to see whether the complexity stays on a different level after a couple of days, or a week or even longer.

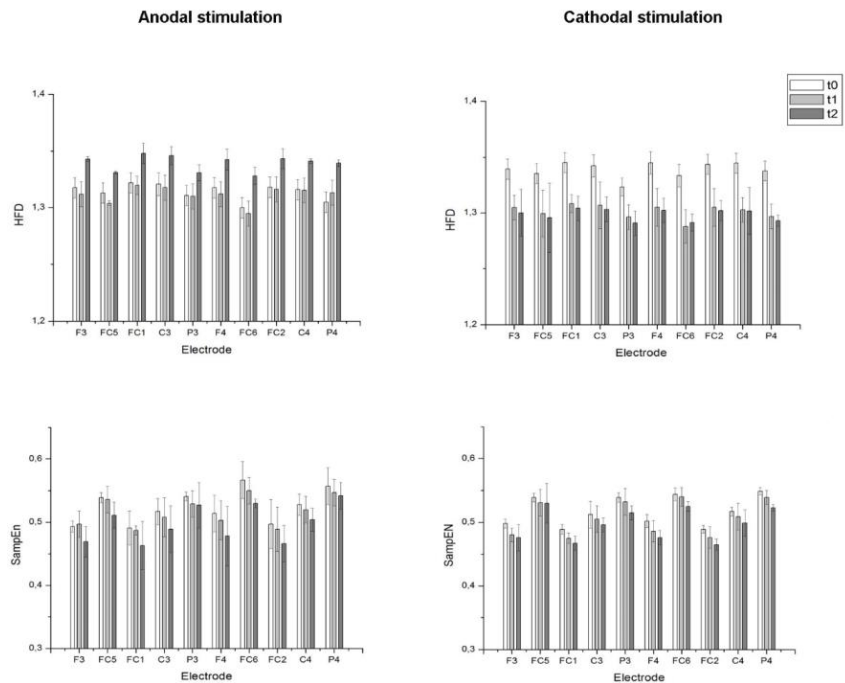

Figure 7: Comparison of complexity changes before and after anodal (left) and cathodal (right) stimulation, represented with two calculated measures Higuchi fractal dimension and sample entropy. The epochs extracted from EEG traces (on ten electrodes) were before the simulation ($t_0$), immediately after the stimulus ($t_1$) and half an hour after the stimulation ($t_2$) (Čukić, 2017).

After all those examples, we can say that both TMS and tDCS are capable of changing the complexity of the system (brain). Although the mechanism and power of their stimulation are different even more than order (power) of the value of the induced field, some similarities can be drawn from our results. If we know that in depression an elevated complexity on cortex is observed, and that majority of researchers agree that this may serve as a biomarker for depression, is it too big a stretch if we think that maybe both rTMS and tDCS can help because they are both diminishing that complexity? If a significant number of people suffering from depression is reporting amelioration of symptoms (let us remember that remission is defined as 'symptoms becoming bearable') they might profit from the ability of electromagnetic stimulation to at least temporarily, decrease the problematic elevated excitability in their DLPFC and elsewhere. We have to stress here, that as well as in ECT, both tDCS and rTMS are requiring maintenance therapy since the benefit reported by patients is lasting for a limited timeframe.

To conclude, further research is needed, on a much larger sample.

**Are rTMS and tDCS capable of changing the complexity of the brain?**

In our previous work, we were interested in complexity changes in the human brain after electromagnetic stimulation (Čukić et al., 2001, 2008, 2013, 2018). The logic behind this is that the complex system (and the brain is one of the most complex dynamical systems that we know) inherently exhibit certain levels of complexity with a possible purpose of accommodation to internal or external changes of parameters which affect the conditions and surroundings of the system is operating in. We have to be aware that we can observe this only indirectly. We are analyzing an EEG as a product of the complex dynamical system (EEG is a composite signal comprising of individual electric signals as a consequence of activation of different neurons and neuronal groups). An electrical signal generated by neural networks when superimposed give us the record from every point we positioned an electrode on the scalp. Of course, we can say that EEG is an electrical representation of brain functioning, but what it is showing us are the voltages picked up from the brain surface (cortex). We can only speculate how that signal represents the deeper structures activities. Despite the extensive use in research and treatment, our knowledge about exact mechanisms by which tDCS is influencing different structures in the brain is limited. Lucia Li and her colleagues examined how transcranial direct current stimulation modulates brain network function (Li et al., 2018). They used MRI for simultaneous recording whit tDCS. From resting state recordings, the main effect of tDCS was to accentuate default mode network (DMN) activation and salience network (SN) deactivation.

In contrast, during task performance, tDCS increased SN activation (Li et al., 2018). In the absence of a task, the main effect of anodal tDCS was more pronounced, whereas cathodal tDCS had a more significant effect during task performance. Cathodal tDCS also accentuated the within

DMN connectivity associated with the performance. There were minimal main effects of stimulation on network connectivity. These results demonstrate that right inferior frontal gyrus tDCS can modulate the activity and functional connectivity of large-scale brain networks involved in cognitive function, in a brain state and polarity dependent manner. Bestmann showed that with sub-threshold stimulation, MRI captured the activation of the auditory system, transversal and superior temporal girys, inferior colliculus and mediate geniculate nucleus, which can be understood as change induced by synaptic transmission (induction of changes of excitability in sensory-motor areas, M1/S1) (Bestmann et al., 2003).

When we apply any artificial influence (like artificially induced electric field in proximity of sensitive neural or muscle tissue) which we know is efficient in inducing action potential (or smaller changes in electrical sense) in the tissue it could be observed through the lens of nonlinear analysis of composite signal (i.e. EEG or EMG)(Čukić, 2006, 2011 and 2013). The levels of complexity induced by this artificial (un-natural) stimulus are changed due to the electrical nature of the phenomena (Čukić, 2011). If we are using TMS we are inducing an artificial electrical field (via Faraday's law) which in turn force electrical charges to move through the tissue (Wagner et al., 2008; Ridding and Rothwell, 2007; Thickbroom, 2007; Bestmann, 2008).

When we try to compare those two stimulations (rTMS and tDCS), the data from the literature are very diverse. We must say that the fundamental biophysical, or physics research followed the advent of TMS just shortly (Paton and Amassian, 1987). Harris and Miniussi elucidated the effects of stimulation in cognitive experimenting focusing on the mechanism of action of induced electrical field on cognitive performances (Harris and Miniussi, 2008), and above mentioned authors explored those details in another, more physiological or biophysical level (like Bestmann, 2008). Both electromagnetic techniques are evolving, and we have to invent better, safer and deeper penetrating noninvasive solutions for depressive disorders.

We would like once again to return to the connection between the electromagnetic stimulation in cases of depression and dynamic of heart rhythm (ECG); we already elaborated on a connection between variability heart rate and response on ECT, or potential forecasting of responders based on nonlinear measures. Researchers who explored the connection of VHR and outcome of therapy concluded that '…low baseline HRV is associated with rapid relapse of depression after ECT. Both high baseline HRV and increasing HRV predict a sustained outcome' (Karpyak et al., 2004). Other researchers in the field also noticed that the low level of VHR corresponds with the severity of disease and that VHR often increases when a patient reaches the remission (Bozkurt et al, 2013). So, without repeating a summary of many physiological complexity studies in this particular task, the conclusion is that the dynamics of the heart is characteristically aberrated in persons diagnosed with bipolar or unipolar depression, and also in burnout syndrome. Lower levels of complexity in heart rhythm are also indicating a potential risk of developing cardiovascular diseases in people with depression.

In comparison with EEG analysis results in depression, we can say that the detection from two physiological signals (EEG and ECG) has the opposite trend, for physiological reasons. In EEG studies, an elevated complexity in depressive patients can be observed and measured (and used as potential neuromarker). In ECG studies, a lower level of nonlinear measures (as well as classical ones) and generally lower complexity can be observed in depressives. However, both of them can be used as useful biomarkers to early detect and monitor the developments in every single case. The difference is that recording of EEG requires a visit to the clinic, while the recording of ECG is already possible with portable monitoring devices. A similar parallel exists between the utilization of rTMS and tDCS; the latter can be used in primary care and even at home, which is a much more accessible option to patients and their families. Like DST was used to differentiate melancholic depression from other less severe depression forms (Shorter and Fink, 2010), so fractal and nonlinear measures of EEG (or ECG) can serve as useful biomarkers necessary for the improvement of accuracy of clinical diagnostic and treatment.

Let us return for a moment to one of the results from our 2016 pilot study; we showed that there is a difference (which can be measured) between the EEG complexity of patients diagnosed with depression who are in acute episode and those who already are in remission. Counterintuitively, those who remitted, exhibited higher levels of complexity, in comparison with participants who were still in exacerbation (Čukić et al., 2018). This phenomenon still needs further exploration. We are currently working on collecting more data, because we believe that even an opposite trend could be present if we screen patients later, after the remission (after possible neural reorganization). Some results from studies dealing with the effect of mindfulness-based cognitive therapy (MBCT) showed that during eight weeks of treatment (or training), the connection between the insula and ventromedial prefrontal cortex were uncoupled, as a marker of improvement measurable within the fronto-limbic system (Davidson et al., 2004; Slagther et al., 2007). It would be interesting to explore further how applications of different measures of complexity on both electrophysiological signals (EEG or ECG) can improve mental healthcare practice. They could become a valuable tool in decision making for clinicians.

When the interaction between electromagnetic stimulation and medication is concerned, we know much more about that, than about the changes of complexity or irregularity, due to a much richer body of evidence in existing literature. For example, among other guidelines, it is recommended in potential tDCS use (Antal et al., 2017) that specific clinical presentation (for example bipolar depressive disorder) requires some mood stabilizers, and not some of the drugs which might attenuate the anodal effect of tDCS. I will add here one anecdotal example as an illustration of that interaction. In the beginning of our project on Mental Healthcare institution, during our conversation about the possibilities of detecting epileptic foci by utilization of a graph theory method based on high density EEG, a colleague mentioned, that it maybe correspond with the fact that in the last five years many patients who are suffering from depression are actually having excellent reaction on anticonvulsants. Is it possible that they have a focus in their limbic system? We discarded that idea as highly unrealistic, but after the finalizing results of several projects on that topic, I think that they are reacting well because of their elevated excitability, induced by their decreased functional connectivity in frontolimbic system. The same can apply to the effectiveness of rTMS and tDCS in cases of treatment-resisting depression; maybe they are effective because of their already demonstrated ability to decrease the levels of complexity in their systems? We also know that the solution is alas, of a temporary nature. As patients reacting well on electroconvulsive therapy need to have later maintenance ECT, it is also required in other forms of electromagnetic stimulation. Both rTMS and tDCS last for a certain amount of time. They are not capable of resolving the problem, and they can help improve the patient's state for a while. The effect is also individual; whatever mode of stimulation we consider, we see that intricate technical details (like the position of electrodes, the physical characteristics of a stimulator), or individual differences (like the thickness of a cortex, or a bone, or different conductivity of a tissue) play their role in final effect of the treatment. Another advantage of the therapeutic use of electromagnetic stimulation is, that opposite to any medication they do not have side effects; or at least they certainly cannot be (or are so far away of being) life-threatening.

Hence, this kind of detection (by utilization of nonlinear measures of electrophysiological signals) are, according to our opinion, one of the prerequisites of a personal medicine in mood disorders.

## Conclusions

Asgdsfg

## Acknowledgements

We want to express our gratitude to our colleagues from Italy Professor Carlo Miniussi, Debora Brignani, Ph.D. and Maria Concetta Pelliciari, Ph.D. for sharing their EEG recordings (data are from the Pellicciari et al., 2013. paper) with us and for all valuable discussions.